\documentclass[11pt]{article}
\usepackage{moriond,epsfig}

\def\be{\begin{equation}}
\def\ee{\end{equation}}
\def\bea{\begin{eqnarray}}
\def\eea{\end{eqnarray}}

\begin{document}
\vspace*{4cm}
\title{THE DEPENDENCE OF GALAXY TYPE ON HOST HALO MASS}

\author{Simone M. Weinmann$^1$, Frank C. van den Bosch$^2$, Xiaohu Yang$^3$,
H.J. Mo$^4$}

\address{$^1$Institute for Theoretical Physics, University of Zurich,
           CH-8057 Zurich, Switzerland\\
         $^2$Max-Planck Institute for Astromomy, D-69117 Heidelberg, Germany\\
         $^3$Shanghai Astronomomical Observatory; the Partner Group of
           MPA, Shanghai 200030, China\\
         $^4$Department of Astronomy, University of Massachusetts, 
           Amherst MA, 01003-9305, USA}

\maketitle

\abstracts{We  examine  the  relation  between galaxy  properties  and
  environment in the SDSS DR2, quantifying environment in terms of the
  mass of the host halo, which  is obtained with a new iterative group
  finder.   We find  that  galaxy type  fractions  scale strongly  and
  smoothly with  halo mass, but, at  fixed mass, not  with luminosity. 
  We compare these findings  with the semi-analytical galaxy formation
  model of  Croton et al.  (2006).   The discrepancies we  find can be
  explained with an oversimplified  implementation of strangulation,
  the neglect  of tidal stripping, and shortcomings  in the treatments
  of dust extinction and/or AGN feedback.}

\section{Introduction}
\label{sec:intro}

It is well established that  there are two different types of galaxies
in the local universe: early type galaxies which have red colours, low
star formation  rates and high  central concentrations, and  late type
galaxies which have blue colours,  are actively forming stars and have
relatively low  central concentrations  (e.g.  Strateva et  al.  2001,
Blanton  et al.   2003). These  two distinct  classes of  galaxies are
found in different environments:  The abundance of early type galaxies
increases  with density  (Dressler  1980) and  towards  the center  of
galaxy groups and clusters (Whitmore  et al. 1993), while the opposite
applies  to late  type galaxies.   

Numerous studies  have reported that galaxy  properties only correlate
with  environment  above  a  characteristic surface  density,  roughly
corresponding to  that at the perimeter  of a cluster  or group (e.g.
Lewis  et   al.   2002).   Consequently,  it  has   been  argued  that
group-specific  processes  play  a  dominant role  in  establishing  a
bimodal  distribution   of  galaxies.   However,   all  these  studies
parameterize  `environment' through  the projected  number  density of
galaxies  above  a  given  magnitude  limit.   Typically  this  number
density,  indicated by  $\Sigma_n$,  is measured  using the  projected
distance to  the $n$th nearest  neighbor.  However, {\it  the physical
meaning of $\Sigma_n$ itself depends on the environment}: in clusters,
where  the number  of galaxies  is  much larger  than $n$,  $\Sigma_n$
measures a {\it local} number  density, which is a sub-property of the
cluster.  However, in low-density environments, which are populated by
haloes which  typically contain only  one or two  galaxies, $\Sigma_n$
measures  a  much  more  global  density, covering  a  scale  that  is
considerably larger than  the halo in which the  galaxy resides.  This
ambiguity severely complicates a  proper interpretation of the various
correlations  between environment and  galaxy properties.   Clearly, a
less ambiguous and more physical  environment indicator is the mass of
the  halo  in   which  the  galaxy  resides.  Here   we  describe  the
construction of a SDSS group catalogue (Section \ref{sec:meth}), which
we use to  analyze how galaxy type depends on  host halo mass (Section
\ref{sec:res}).  In  Section \ref{sec:comp} we  compare these findings
to  a recent  semi-analytical model  of galaxy  formation in  order to
constrain mechanisms that cause a quenching of star formation.

\section{Methodology}
\label{sec:meth} 

\subsection{The Data Set}

The data  used in  our analysis  is taken from  the Sloan  Digital Sky
Survey  DR2 (SDSS).   In particular,  we use  the New  York University
Value-Added Galaxy  Catalogue (NYU-VAGC,  Blanton et al.   2005). From
this  catalogue we select  all galaxies  with an  extinction corrected
apparent  magnitude $r<18$,  $0.01  < z  <  0.2$ and  with a  redshift
completeness  of $c  > 0.7$.   In addition,  we use  estimates  of the
specific star  formation rates  (SSFR) obtained by  Brinchmann et  al. 
(2004).   Throughout   our  analysis  we   use  Petrosian  magnitudes,
$k$-corrected to $z=0.1$,  and we adopt a Hubble  constant of $100\, h
\, {\rm km} {\rm s}^{-1} {\rm Mpc}^{-1}$.

\subsection{Defining Galaxy Types}
\label{subsec:type}

\begin{figure}
\psfig{figure=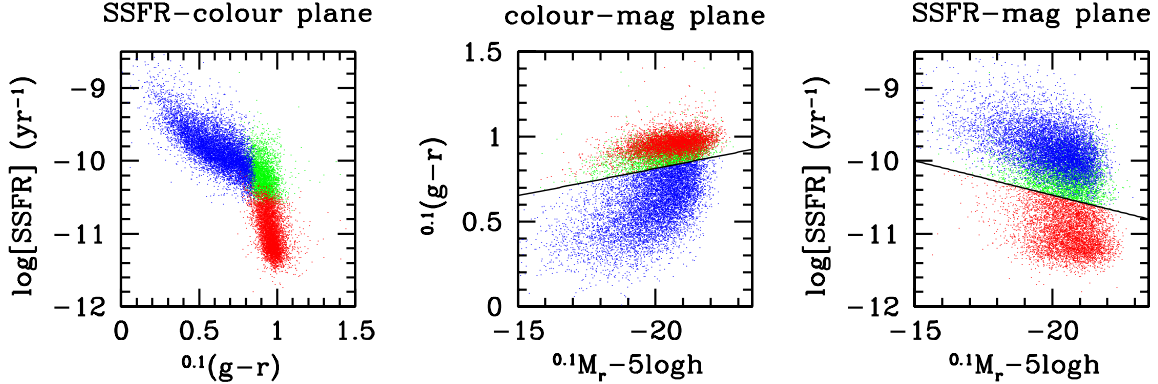,height=2in}
\caption{The location  of the galaxies used  as an input  to our group
  finder on the plane of colour and specific star formation rate (left
  hand panel), colour and absolute r-band magnitude (middle panel) and
  specific  star formation  rate and  absolute r-band  magnitude. Blue
  dots denote 'late type galaxies', red dots 'early type galaxies' and
  green dotes 'intermediate type galaxies' (see text for definition).}
\label{fig:col}
\end{figure}

Since galaxies follow a  relatively narrow relation between colour and
SSFR, consisting of two overlapping branches (Fig.~\ref{fig:col}, left
hand  panel), we  classify galaxies  using both  their SSFR  and their
$^{0.1}(g-r)$-colour. Using  the magnitude dependencies  of the color-
and SSFR  bimodalities (shown as solid  lines in the  middle and right
hand panels of Fig.~\ref{fig:col}),  we split the galaxy population in
three  main types:  `late-type' galaxies,  defined as  being  blue and
active (blue dots, 48 percent of all galaxies), `early-type' galaxies,
defined  as  being  red and  passive  (red  dots,  31 percent  of  all
galaxies), and `intermediate-type' galaxies,  defined as being red and
active (green  dots, 20 percent  of all galaxies).  The  remaining one
percent are  blue and passive,  and we do  not consider this  class in
what follows.

\subsection{Identification of Galaxy Groups}
\label{subsec:group}

Galaxy  groups are  identified  using a  new,  iterative group  finder
developed by  Yang et al.  (2005),  which is optimized  to group those
galaxies  together that  belong to  the  same dark  matter halo.   Its
performance has been  tested in terms of completeness  of true members
and contamination by interlopers,  using detailed mock galaxy redshift
surveys.  The  average completeness of individual groups  is $\sim 90$
percent and they contain only  $\sim 20$ percent interlopers, which is
a  significant  improvement  with  respect to  the  more  conventional
Friends-of-Friends group finders.

Group masses  are estimated by  matching the number density  of groups
sorted according to  their total luminosity, with that  of dark matter
haloes sorted according to their mass.  This makes the assumption that
there is a one-to-one relation between total group luminosity and halo
mass,  and  requires  the  halo  mass  function,  which  is  cosmology
dependent. Nevertheless, detailed tests have shown that this method of
assigning masses  to groups is  more accurate than using  the velocity
dispersion of  their member galaxies,  especially for low  mass groups
(see Weinmann et al.  2006a for details).

Application  of  our group  finder  to  the  sample of  SDSS  galaxies
described above  yields a  group catalogue of  53,229 systems  with an
estimated mass, containing a total of 92,315 galaxies. In what follows
we  refer to  the  brightest galaxy  in  each group  as the  `central'
galaxy, while all others are termed `satellites'.

\section{Results: The Galaxy Type - Halo Mass Relation}
\label{sec:res}

\begin{figure}
\begin{center}
\psfig{figure=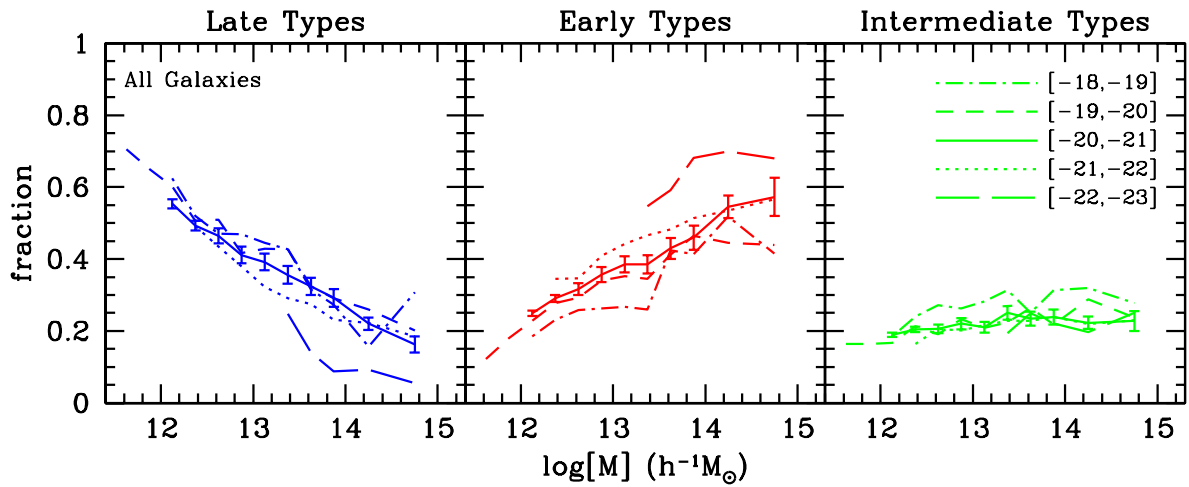,height=2.5in}
\caption{The fraction  of late type galaxies, early  type galaxies and
  intermediate type  galaxies as a  function of group  mass. Different
  line styles denote different absolute magnitude bins as indicated.}
\label{fig:res}
\end{center}
\end{figure}

Fig. \ref{fig:res} plots the fractions  of late types, early types and
intermediate types  as function of  group mass. Results are  shown for
five different bins in  absolute magnitude (different line-styles), as
indicated. As  expected, the fraction of late  (early) types decreases
(increases) with  increasing halo mass, basically  reflecting the well
known morphology-density  relation. Note, however, that  at fixed halo
mass there is virtually  no luminosity dependence, indicating that the
mass  of the  host halo  is much  more important  for  determining the
galaxy type than  the luminosity of the galaxy  itself. This indicates
that the well-known relation between  galaxy type and luminosity (e.g. 
Baldry  et al.  2004) is  mainly a  reflection of  the fact  that more
luminous galaxies tend to live in more massive haloes.  

Note  also   that  the  mass-dependence  is   remarkably  smooth.   In
particular, there is  no indication of any characteristic  mass scale. 
This is  contrary to  several previous studies,  who have  argued that
group-specific  processes  play  a  dominant role  in  establishing  a
bimodal    distribution     of    galaxies.     As     discussed    in
Section~\ref{sec:intro}, we  belief that this owes to  the ambiguity of
the density indicator used in those previous studies.

Finally, we  find no dependence  of the intermediate type  fraction on
either  group  mass  or   galaxy  luminosity.   This  has  interesting
implications. If the intermediate types  consist of a mix of early and
late types, their fractional contribution  must be close to 50 percent
at all luminosities, and in haloes of all masses, which seems somewhat
contrived.   More  likely,  the  intermediates are  actual  transition
objects, transiting  from the `blue  cloud' to the `red  sequence', or
vice versa.  In this case, their  observed fractions can  teach us about
the  transition probabilities  and durations.  It remains  to  be seen
whether these are consistent with predictions of galaxy formation models. 

\section{Comparison with a Semi-Analytic Model}
\label{sec:comp}
  
A successful  reproduction of the  galaxy luminosity function,  and of
the bimodality  in the galaxy distribution, requires  a mechanism that
can  truncate star  formation in  massive haloes.   Current  models of
galaxy   formation   consider    two   such   truncation   mechanisms:
strangulation,  which acts  on satellite  galaxies, and  AGN feedback,
which predominantly  affects central galaxies.   The efficiencies with
which these two  processes operate set the blue  fraction of galaxies,
$f_{\rm blue}(L,M)$,  as function of galaxy luminosity,  $L$, and halo
mass, $M$.  We  now compare the blue fractions  obtained from our SDSS
group catalogue with those  obtained from the semi-analytical model of
Croton et al.  2006 (hereafter  C06). This model includes 'radio mode'
AGN feedback  and is  very successful in  reproducing a wide  range of
global properties of the galaxy distribution.

In order to carry out a fair comparison, we construct a mock SDSS from
the SAM, to which we apply  our group finder. The resulting mock group
catalogue,  hereafter `SAM-GC',  can  be compared  to  our SDSS  group
catalogue on a one-to-one basis. Contrary to the analysis above, where
we classified galaxies based on  both their color and their SSFR, here
we only split  galaxies in red and blue  sub-classes, using the colour
cut shown in the middle panel of Fig.~\ref{fig:col}.

\begin{center}
\begin{figure}
\psfig{figure=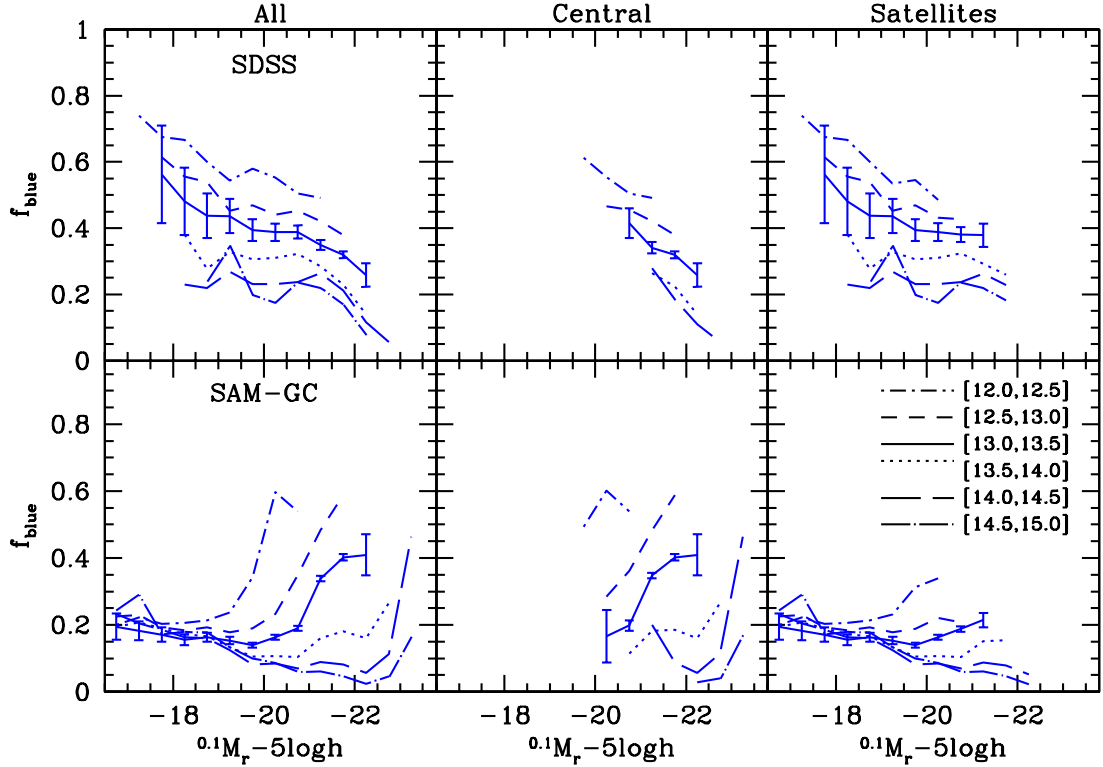,height=4in}
\caption{The  fraction  of  blue  galaxies  as a  function  of  galaxy
  luminosity in  the SDSS (top  panel) and the semi-analytic  model of
  C06  (bottom panel).  Results  are shown  for 6  halo mass  bins, as
  indicated,  whereby the  values  in square  brackets indicate  ${\rm
    log}(M/h^{-1}M_{\odot})$.}
\label{fig:sa1}
\end{figure}
\end{center}

We split  the SDSS  and SAM group  catalogues in six  logarithmic mass
bins and determine how the blue fractions in each of these bins depend
on luminosity.  The results for  the SDSS group catalogue are shown in
the  upper panels  of Fig.~\ref{fig:sa1}.   The upper  left-hand panel
shows the result for all galaxies (centrals plus satellites).  In each
mass bin, the blue  fraction decreases with increasing luminosity, but
only very mildly.  At fixed luminosity, however, there is a clear mass
dependence,  with the  blue fraction  decreasing with  increasing halo
mass. This  is consistent  with Fig.~\ref{fig:res} and  indicates that
the colour of a galaxy is  more strongly determined by the mass of the
halo in which  it resides than by its own  luminosity.  The middle and
right hand panels in the upper row show the blue fractions of centrals
and satellites in the SDSS.

Comparing  these blue fractions  with those  obtained from  the SAM-GC
(lower  panels)  one notices  two  problems:  First  of all,  the  SAM
predicts blue  satellite fractions that  are much too  low, especially
for  low   mass  haloes.   This  suggests   that  `strangulation',  as
incorporated  in the  SAM, is  much too  efficient.  In  virtually all
semi-analytical models,  strangulation is included and  modeled in the
same way: as  soon as a galaxy becomes a satellite  galaxy its hot gas
reservoir is `stripped' (see Larson et al. 1980).  Consequently, after
a delay  time in which  the new satellite  consumes its cold  gas, its
star formation is truncated.   The results presented here suggest that
this formulation is too crude. In particular, strangulation is modeled
without any explicit halo mass dependence, which explains why the blue
fraction of satellite galaxies in  the SAM is virtually independent of
halo  mass.   In  the  SDSS,  however,  the  blue  satellite  fraction
decreases with  increasing halo mass, suggesting a  clear mass scaling
of the  strangulation efficiency.  

The second  problem with  the SAM  is that it  predicts that  the blue
fraction  of centrals  increases with  luminosity in  haloes  of fixed
mass, opposite  to what is seen in  the data. As shown  in Weinmann et
al. (2006b),  a related problem is  that the fraction  of bright, blue
central galaxies is too high in  the SAM. Both of these problems could
be due  to an  oversimplified treatment of  dust extinction  (see C06)
and/or a problem with the treatment of AGN feedback in the model. 

Finally,  Fig.  4  shows  the conditional  luminosity function  (CLF),
which specifies the  average number of galaxies of  a given luminosity
per group. Results  are shown for four group  masses (indicated at the
top)  and for both  the SDSS  and SAM  group catalogues.   The overall
agreement  is remarkable,  although in  the highest  mass bin  the SAM
significantly  overpredicts the  abundance of  faint  satellites. This
might be due to the neglect of tidal stripping in the SAM, which could
destroy, or at  least dim faint satellite galaxies  to the extent that
they are not detectable anymore.

\begin{centering}
\begin{figure}
\psfig{figure=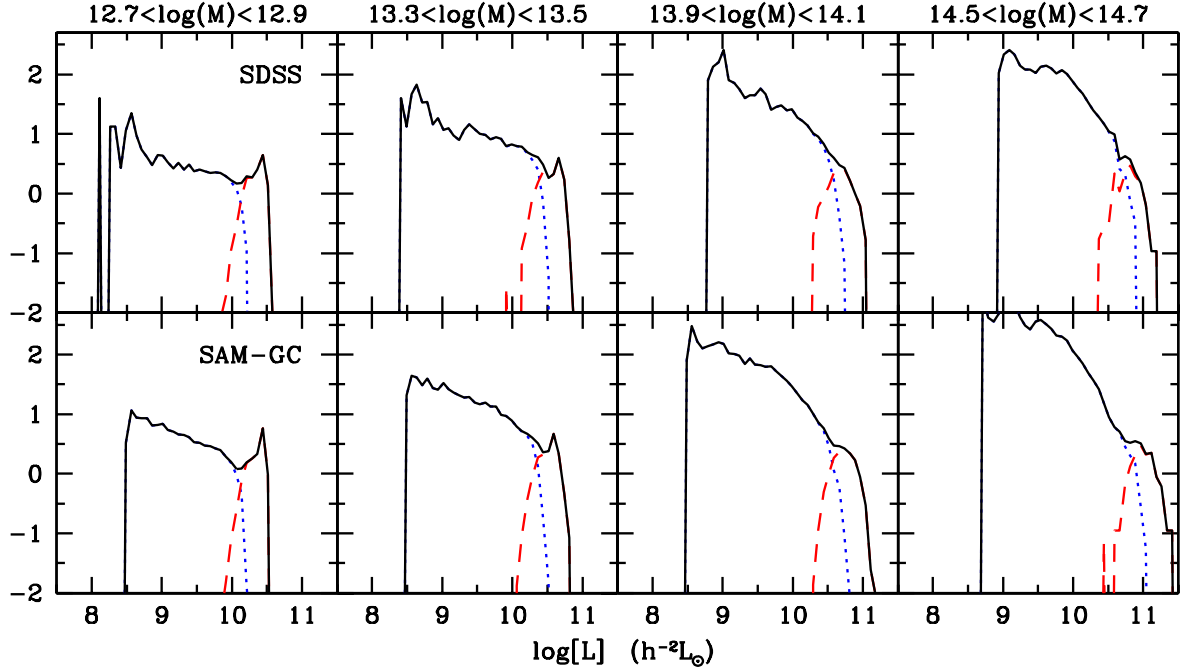,height=3.5in}
\caption{The conditional luminosity  function (CLF) for four different
  halo mass  bins (indicated in the  top). Top and  bottom panels show
  results  for  SDSS  and   SAM-GC,  respectively.  The  agreement  is
  remarkably good, but note that the SAM overpredicts the abundance of 
  faint satellites in the most massive haloes.}
\label{fig:sa2}
\end{figure}
\end{centering}

\section{Summary}

Using a  group catalogue based  on the SDSS  DR2, we have  derived the
relations  between galaxy type  fractions and  group masses.   We find
that the fraction of late  type galaxies decreases smoothly with group
mass and  is nearly  independent of galaxy  luminosity at  fixed group
mass. We also find that  the fraction of intermediate type galaxies is
independent  of  both  group  mass and  luminosity.   These  relations
provide  a useful  test-bed for  models  of galaxy  formation.  As  an
example, we have  compared our findings to the  semi-analytic model of
Croton  et al (2006),  which successfully  matches many  {\it global}
properties of the local galaxy  population.  However, when it comes to
the properties as  function of halo mass, we  find several significant
discrepancies  between model and  data.  First  and foremost,  the SAM
predicts too  many red  satellites, indicating that  the strangulation
treatment  in the SAM  is far  too efficient.   Second, the  number of
faint  satellites in  the SAM  is overestimated,  probably due  to the
neglect of  tidal stripping  in the model.   Finally, the  fraction of
bright, blue central  galaxies is too high in the  SAM, which could be
due  to a  too simple  dust model  or also  due to  problems  with the
implementation  of AGN  feedback.   Clearly, the  data presented  here
regarding  the relations  between galaxy  type and  halo  mass provide
important constraints  on galaxy formation models. We  hope these data
will  prove useful  to  test  and calibrate  future  models of  galaxy
formation and in particular to discriminate between various models for
AGN  feedback and  other  star formation  truncation mechanisms.  More
details,  and  the actual  data,  can  be found  in  Weinmann  et al.  
(2006a,b).


\section*{References}
\begin{thebibliography}{99}

\bibitem{ba}      I.K.     Baldry,      K.Glazebrook,     J.Brinkmann,
  \v{Z}.  Ivezi\'{c},  R.H. Lupton,  R.C.  Nichol,  A.S. Szalay,  ApJ,
  600,681 (2004)

\bibitem{bl} M.R. Blanton {\it et al}, ApJ, 594, 186 (2003)

\bibitem{bl5} M.R. Blanton {\it et al}, AJ, 129, 2562 (2005)

\bibitem{br}  J.  Brinchmann,  S.  Charlot, S.D.M  White,  C.Tremonti,
  G. Kauffmann, T. Heckman, J. Brinkmann, MNRAS, 353, 713 (2004)

\bibitem{cr} D. Croton {\it et al}, MNRAS, 365, 11 (2006) (C06)

\bibitem{dr} A. Dressler, ApJ, 236, 351 (1980)

\bibitem{la} R.B. Larson, B.M. Tinsley, C.N. Caldwell, 1980, ApJ, 237,
692

\bibitem{le} I. Lewis {\it et al}, MNRAS, 334, 673 (2002)

\bibitem{st} I. Strateva {\it et al}, ApJ, 122, 1861 (2001)

\bibitem{we} S.M Weinmann, F.C. van den Bosch, X. Yang, H.J.  Mo,
  MNRAS, 366, 2 (2006a)

\bibitem{we2} S.M. Weinmann, F.C. van den Bosch, X. Yang, H.J. Mo,
D.J. Croton, B. Moore(2006b), preprint (astro-ph/0606458)

\bibitem{wh}  B.C. Whitmore,  D.M Gilmore,  C.  Jones,  ApJ,  407, 489
  (1993)

\bibitem{ya} X. Yang, H.J  Mo, F.C van  den Bosch, Y.P.  Jing, MNRAS,
356, 1293 (2005)

\end{thebibliography}
\end{document}